\documentclass{article}
\usepackage[utf8]{inputenc}
\usepackage[affil-it]{authblk}
\usepackage{tikz}
\usetikzlibrary{arrows,shapes,positioning,shadows,trees}
\tikzset{
  basic/.style  = {draw, text width=2cm, drop shadow, font=\sffamily, rectangle},
  root/.style   = {basic, rounded corners=2pt, thin, align=center,
                   fill=green!30},
  level 2/.style = {basic, rounded corners=6pt, thin,align=center, fill=green!60,
                   text width=8em},
  level 3/.style = {basic, thin, align=left, fill=pink!60, text width=6.5em}
}

\usepackage{hyperref}
\usepackage{graphicx}
\graphicspath{{img/}}
\usepackage{tabularx}
\newcommand{\tableheadline}[1]{\multicolumn{1}{c}{#1}}
\usepackage{dirtree}
\usepackage{rotating}
\usepackage{booktabs}

\usepackage[frozencache]{minted}
\definecolor{dhscodebg}{rgb}{0.95,0.95,0.95} % listing background
\DeclareUnicodeCharacter{2212}{-}

\usepackage{xspace}

\usepackage{csquotes}

\newcommand{\mypddl}{\textsc{myPddl}\xspace}
\newcommand{\mypddlclojure}{\textsc{myPddl-clojure}\xspace}
\newcommand{\mypddlsyntax}{\textsc{myPddl-syntax}\xspace}
\newcommand{\mypddldiagram}{\textsc{myPddl-diagram}\xspace}
\newcommand{\mypddlnew}{\textsc{myPddl-new}\xspace}
\newcommand{\mypddlide}{\textsc{myPddl-ide}\xspace}
\newcommand{\mypddlsnippet}{\textsc{myPddl-snippet}\xspace}
\newcommand{\mypddldistance}{\textsc{myPddl-distance}\xspace}
\newcommand{\mypddlplan}{\textsc{myPddl-plan}\xspace}
\newcommand{\pddlstudio}{\textsc{pddl studio}\xspace}
\newcommand{\epd}{editor.planning.domains\xspace}
\newcommand{\itsimple}{\textsc{itSimple}\xspace}
\newcommand{\pddlmode}{\textsc{pddl}-mode\xspace}
\newcommand{\vscode}{vscode-\textsc{pddl}\xspace}
\newcommand{\pddl}{\textsc{pddl}\xspace}
\newcommand{\uml}{\textsc{uml}\xspace}

\newcommand{\sublimetext}{Sublime Text\xspace}

\begin{document}
\title{MyPDDL: Tools for efficiently creating PDDL domains and
  problems}
\renewcommand*{\Affilfont}{\small}
\author[1]{Volker Strobel%
  \thanks{\texttt{vstrobel@ulb.ac.be} (corresponding author)}}
\author[2]{Alexandra Kirsch}
\affil[1]{IRIDIA, Universit\'e Libre de Bruxelles, Belgium}
\affil[2]{Independent Scientist}
\date{}

%\author{Volker Strobel \and
%Alexandra Kirsch}
%
%\authorrunning{V. Strobel and A. Kirsch}
%\institute{IRIDIA, Universit\'e Libre de Bruxelles, Belgium \and
%Independent Scientist\\
%\email{vstrobel@ulb.ac.be}}
%
\maketitle              % typeset the header of the contribution
\begin{abstract}
  The Planning Domain Definition Language (PDDL) is the
  state-of-the-art language for specifying planning problems in
  artificial intelligence research. Writing and maintaining these
  planning problems, however, can be time-consuming and error
  prone. To address this issue, we present myPDDL---a modular toolkit
  for developing and manipulating PDDL domains and problems. To
  evaluate myPDDL, we compare its features to existing knowledge
  engineering tools for PDDL. In a user test, we additionally assess
  two of its modules, namely the syntax highlighting feature and the
  type diagram generator. The users of syntax highlighting detected
  36\,\% more errors than non-users in an erroneous domain file. The
  average time on task for questions on a PDDL type hierarchy was
  reduced by 48\,\% when making the type diagram generator
  available. This implies that myPDDL can support knowledge engineers
  well in the PDDL design and analysis process.
  
%\keywords{PDDL \and Planning \and Knowledge Engineering}
\end{abstract}

\section{Introduction}
\label{sec:introduction}

Being a key aspect of artificial intelligence (AI), \emph{planning} is
concerned with devising a sequence of actions to achieve a desired
goal~\cite{helmert2008understanding}. AI planning has made remarkable
progress in solving planning problems in large state spaces that would
be impossible for humans to handle. The International Planning
Competition\footnote{\url{http://www.icaps-conference.org/index.php/Main/Competitions/}}
has led to a number of open source planners that are ready to be used
by practitioners and researchers outside the AI planning field.

However, the effectiveness of planning largely depends on the quality
of the problem formalization~\cite{shah2013knowledge}. \pddl
(\emph{Planning Domain Definition Language})~\cite{mcdermott1998pddl}
is the de facto standard for the description of planning
tasks~\cite{ilghami2005extension}. It divides the description of a
planning task into a domain model and problem descriptions: the
description of a household with its objects and locations would be a
domain, with possible tasks such as making breakfast, cleaning the
windows, or changing a light bulb. Someone with a non-planning
background, for example, from robotics, has to get used to the \pddl
syntax and its possibilities to describe the world. She also has to
keep track of the facts in the domain model and the different planning
tasks. While automated planning can save vast amounts of time to find
a valid solution, creating the planning task specifications is a
complex, error-prone, and cumbersome task. An ill-defined problem is
often the reason for finding suboptimal plans or no plan at all.  The
household domain in this example comes with another challenge. For
many tasks the distances between objects or other numerical input may
be necessary. \pddl is by its very nature as a planning language
designed for symbolic specifications. To use numerical data
efficiently, it must often be preprocessed.

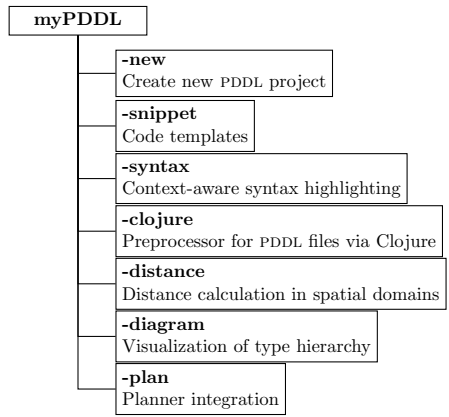
\begin{figure}[H]
  \centering
  \tikzstyle{every node}=[draw=black,anchor=west,
  align=left,
  scale=0.7]
  \begin{tikzpicture}[%
  grow via three points={one child at (0.5,-0.7) and
    two children at (0.5,-0.7) and (0.5,-1.4)},
  edge from parent path={(\tikzparentnode.south) |- (\tikzchildnode.west)}]
  \node {~~~\textbf{myPDDL}~~~}
  child { node {\textbf{-new}\\Create new \pddl project}}
  child { node {\textbf{-snippet}\\Code templates}}
  child { node {\textbf{-syntax}\\Context-aware syntax highlighting}}
  child { node {\textbf{-clojure}\\Preprocessor for \pddl files via Clojure}}
  child { node {\textbf{-distance}\\Distance calculation in spatial domains}}
  child { node {\textbf{-diagram}\\Visualization of type hierarchy}}
  child { node {\textbf{-plan}\\Planner integration}};
  \end{tikzpicture}
    \caption{\mypddl is a highly customizable and extensible modular
    system, designed for supporting knowledge engineers in the process
    of writing, analyzing and expanding \pddl files and thereby
    promoting the collaboration between knowledge engineers and the
    use of \pddl in real-world applications. It consists of the parts
    shown in the figure.}
  \label{fig:mypddl-overview}  
\end{figure}

In this chapter, we describe
\mypddl~(Figure~\ref{fig:mypddl-overview}), a knowledge engineering
toolkit that supports knowledge engineers in the entire design cycle
of specifying planning tasks without having to become an expert in AI
planning. In the initial stages, it allows for the creation of
structured \pddl projects that should encourage a disciplined design
process. With the help of snippets, that is, code templates, often
used syntactic constructs can be inserted into \pddl files. A syntax
highlighting feature that speeds up the error-detection supports
intermediate stages. Understanding the textual representation of
complex type hierarchies in domain files can be confusing, so an
additional tool enables their visualization. \pddl's limited modeling
capabilities were bypassed by developing an interface that converts
\pddl code into code of the functional programming language Clojure
\cite{hickey2008clojure} and vice versa.  Within this project, the
interface was employed for a feature that calculates distances between
objects specified in a problem model, but the interface provides
numerous other possibilities and could also be used to further
automate the modeling process. A basic planner integration allows for
quickly running a desired planner. All of the features were integrated
into the customizable and extensible Sublime
Text\footnote{\url{http://www.sublimetext.com/}} editor.

Since the main aim in the development of the toolkit was for it to be
easy to use and maintain, it is evaluated with regard to these
criteria. Another aim was to make planning more accessible in
real-life tasks and to enable inexperienced users to get started with
planning problems. Therefore, \mypddl's usability was assessed by
means of a user test with eight subjects that had no prior experience
with AI planning. The results show that \mypddl facilitates both
error-detection and the understanding of a given domain.

This chapter is an extended version of work published in a previous
paper~\cite{StrKir2014:aai}. The remainder of this chapter is
structured as follows. Section~\ref{sec:related-work} compares \pddl
knowledge engineering tools to lay a foundation for
\mypddl. Section~\ref{sec:mypddl} describes the different modules of
\mypddl and their design principles. Section~\ref{sec:valid-eval}
evaluates \mypddl via a user test. Section~\ref{sec:conclusion}
concludes the chapter and outlines future work.

\section{Related Work}
\label{sec:related-work}

This section introduces, compares, and discusses knowledge engineering
tools that allow text-based editing of \textsc{pddl} files to set the
stage for \mypddl (Table~\ref{tab:comparison}).

\pddlstudio \cite{plch2012inspect} is an IDE (\emph{integrated
  development environment}) for creating and managing \textsc{pddl}
projects, that is, a collection of \textsc{pddl} files. Its main
features are syntax highlighting, error detection, context sensitive
code completion, code folding, project management, and planner
integration. Many of these features are based on a parser, which
continuously analyzes the code and divides it into syntactic
elements. These elements and the way in which they relate to each
other can then be identified. The syntax highlighter is a tool that
colors constructs according to their syntactical meaning within the
code. In the case of \pddlstudio, it colors names, variables, errors,
keywords, predicates, types, and brackets each in a different
customizable color.  \pddlstudio's error detection can recognize both
syntactic (missing keywords, parentheses, etc.) and semantic (wrong
type of predicate parameters, misspelled predicates, etc.)
errors. This means that \pddlstudio can detect errors based on a
mismatch between domain and problem file in real time. The code
completion feature offers recommendations for standard \textsc{pddl}
constructs as well as for previously used terms. Code folding allows
the knowledge engineer to hide currently not needed code blocks. In
this case only the first line of the block is displayed. Lastly, a
command-line interface allows the integration of planners in order to
run and compare different planning software.

Unlike \pddlstudio, which provides a text based editor for \pddl, the
\itsimple \cite{vaquero2005itsimple} editor has, as its main feature,
a graphical approach that allows designing planning tasks in an
object-oriented approach using \uml (\emph{Unified Modeling
  Language}). In the process leading up to \itsimple,
\textsc{uml.p}---\textsc{uml} in a Planning Approach---was proposed,
which is a \textsc{uml} variant specifically designed for modeling
planning domains and problems \cite{vaquero2006use}.

The main purpose of \itsimple is supporting knowledge engineers in the
initial stages of the design phase by making tools available that help
transform the informality of the real world to formal specifications
of domain models. The professed aim of the project is to provide a
means to a \enquote{disciplined process of elicitation, organization
  and analysis of requirements} \cite{vaquero2005itsimple}. However,
subsequent design stages are also supported. Once domain and problem
models have been created, \textsc{pddl} representations can be
generated from the \textsc{uml.p} diagrams, edited, and then used as
input to a number of different integrated planning systems.

With \itsimple, it is possible to directly input the domains and
problems into a planner and to inspect the output from the planning
system using the built-in plan analysis. This consists of a plan
visualization, which shows the interaction between the plan and the
domain by highlighting every change caused by an action. \itsimple's
modeling workflow is unidirectional as changes in the \pddl domain do
not affect the \uml model and \uml models have to be modeled manually,
meaning that they cannot by generated using \pddl.

Starting in version 4.0, \itsimple expanded its features to allow the
creation of \pddl projects from scratch, that is, without the \uml to
\pddl translation process \cite{vaquero2012itsimple4}. So far, a basic
syntax highlighting feature recognizes \pddl keywords, variables, and
comments. \itsimple also provides templates for \pddl constructs, such
as requirement specifications, predicates, actions, initial state, and
goal definitions.

\pddlmode\footnote{\url{http://rakaposhi.eas.asu.edu/planning-
    list-mailarchive/msg00085.html}} for Emacs builds on the
sophisticated features of the widely used Emacs editor and uses its
extensibility and customizability. \pddlmode provides syntax
highlighting by way of basic pattern matching of keywords, variables,
and comments. Additional features are automatic indentation and code
completion as well as bracket matching. Code snippets for the creation
of domains, problems, and actions are also available. Finally,
\textsc{pddl}-mode keeps track of action and problem declarations by
adding them to a menu and thus intending to allow for easy and fast
code navigation.

\pddlmode for Emacs supports \textsc{pddl} versions up to
2.2, which includes derived predicates and timed initial predicates
\cite{edelkamp2004pddl2}, but does not recognize later features like
object-fluents.

The online tool \epd allows for editing \pddl files in a web
browser. Its features comprise syntax highlighting, code folding,
\pddl-specific auto-completion, and multi-tab support. The editor is
part of the Planning-Domains\footnote{\url{http://planning.domains/}}
initiative which aims at providing three pillars to the planning
community: (1) an API to access existing \pddl domains and problems;
(2) a planner-in-the-cloud service which be be accessed via a RESTful
API; and (3) an online \pddl editor. The online editor is also
connected to the planner in the cloud.

The \pddl plugin
\vscode\footnote{\url{https://marketplace.visualstudio.com/items?itemName=jan-dolejsi.pddl}}
for the editor VS Code (\emph{Visual Studio Code}) offers a wide range
of editing functions, such as syntax highlighting, code completion,
code folding, and code snippets. It offers a mature planner
integration and plan visualization. Thanks to a \pddl parser
integration, it is possible to detect semantic errors immediately when
they are made.

\subsection{Critical Review}
\label{sec-3-4}

All the above-mentioned tools provide environments for the creation of
\pddl code. Their advantages and disadvantages are reviewed in this
section. At the end of each discussed feature, the approach that was
used in \mypddl is introduced.

\pddlstudio, \itsimple, and \epd for the most part do not build on
existing editors and therefore cannot fall back on refined
implementations of features, such as selection of tab size, defining
custom key shortcuts, customizing the general look and feel, and
bracket matching. In contrast, \vscode and \pddlmode for Emacs are
integrated into mature code editors and can be used in combination
with other plugins. To have both basic editor features and a high
customizability, it was decided to use an existing, extensible text
editor to integrate \mypddl into.

The tools can also be compared in terms of their syntax highlighting
capabilities. In \pddlmode for Emacs (up to \pddl\,2.2) and \epd (up
to \pddl\,3.1), and \vscode (up to \pddl\,3.1) keywords, variables,
and comments are highlighted. However, this is only done via pattern
matching without controlling for context. This means that wherever the
respective terms appear within the code they will get highlighted,
regardless of the syntactical correctness. Different colors can be
chosen by customizing Emacs and Visual Studio Code. \epd provides two
fixed color schemes. \itsimple's syntax highlighting for \pddl 3.1 is,
except for the \pddl version difference, equally as extensive as that
of \pddlmode for Emacs but does not allow for any customization.
\pddlstudio has advanced syntax highlighting that distinguishes all
different \pddl 1.2 constructs depending on the context and allows
knowledge engineers to choose their preferred highlighting colors. One
of the primary objectives of \mypddl is to help users in keeping track
of their \pddl programs. As a means to this end, it was decided to
also implement sophisticated, context-dependent syntax highlighting.

Another useful feature for fast development is the ability to insert
larger code skeletons or snippets. \pddlstudio does not support the
insertion of code snippets. \itsimple features some code templates for
predicates, derived predicates, functions, actions, constraints,
types, comments, requirements, objects, and metrics. However, the
templates are neither customizable nor extensible. \pddlmode for Emacs
provides three larger skeletons: one for domains, one for problems,
and one for actions. Further skeletons could be added. Both \epd and
\vscode provide many code snippets. \mypddl aims to combine the best
of these latter tools and support customizable and extensible snippets
for domains, problems, types, predicates, functions, actions, and
durative actions.

\pddlstudio, \pddlmode for Emacs, and \epd do not provide
visualization options. \itsimple, on the other hand, is based entirely
on visually modeling domains and problems. Therefore, since the first
version, the focus has mainly been on exporting from \textsc{uml.p} to
\pddl and to visualize plans. \mypddl is to reverse this design
approach and enable type diagram visualization of some parts of the
\pddl code. \vscode does not provide domain visualization but is able
to visualize a found plan.

Searching for errors can be one of the most time consuming parts of
the design process. Hence, any tool that is able to help detect errors
faster is of great value to the knowledge engineer. While \pddlmode
for Emacs, \itsimple, and \epd facilitate error detection only by
basic syntax highlighting, both \pddlstudio and \vscode are able to
detect errors via a \pddl parser. In \mypddl, a different approach is
taken and syntactic errors are \emph{not} highlighted by the syntax
highlighting feature, while all correct \pddl code \emph{is}
highlighted.

A major drawback of \pddlstudio and \pddlmode for Emacs especially is
that they are not updated regularly to support the most recent \pddl
versions. \pddlstudio's parser is only able to parse \pddl 1.2, while
the latest \pddl version is 3.1. \pddl has significantly evolved since
\pddl 1.2 and was extended in \pddl 2.1 to include \emph{durative
  actions} to model time dependent behaviors, \emph{numeric fluents}
to model non-binary changes of the world state, and
\emph{plan-metrics} to customize the evaluation of plans. \pddlmode
for Emacs is only compatible with \pddl versions up to 2.2, which
introduced \emph{derived predicates} and \emph{timed initial
  predicates} but does not recognize later features like
\emph{object-fluents}. It follows that the range of functions
specified in the domain file cannot include object-types in addition
to numbers. \itsimple, \epd, \vscode, and \mypddl support the latest
\pddl version.

\pddlstudio falls short of customization options since they are
limited to the choice of font style and color of highlighted \pddl
expressions. Furthermore, \pddlstudio is written as standalone
program, meaning that there are no \pddl-independent extensions. The
same holds true for \itsimple. Since both Emacs and VS Code are
established editors, \pddlmode and \vscode are highly customizable and
extensible. This is the other major reason why it was decided that
\mypddl should be integrated into an existing, extensible, and
customizable text editor. These requirements are met by Sublime Text,
a text editor that offers a wide variety of features and plugins.

All in all, \mypddl must be understood as complementary to the other
existing knowledge engineering tools. \mypddl is distributed as a
package for Sublime Text and provides context-aware syntax
highlighting, code snippets, syntactic error detection, and type
diagram visualization. Additionally, it allows for the automation of
modeling tasks due to an interface with Clojure that supports the
conversion of \textsc{pddl} code into Clojure code and vice
versa. Therefore, \mypddl is intended to support both the initial
design process of creating domains with code snippets, syntax
highlighting and the Clojure interface and the later step of checking
the validity of existing domains and problems with the type diagram
generator. Lastly, the visualization capabilities of \mypddl are meant
to facilitate collaboration among knowledge engineers.

\begin{sidewaystable}[]
\centering
\scriptsize
\caption[Comparison of knowledge engineeringtools]{\label{tool-comp}Comparison of knowledge engineering tools and their features.}
\begin{tabularx}{\textwidth}{lX|llllll}
  \tableheadline{Feature}             & \tableheadline{Function}                              & \tableheadline{\pddlstudio} & \tableheadline{\itsimple} & \tableheadline{\pddlmode} & \tableheadline{planning.domains} & \tableheadline{\vscode} & \tableheadline{\mypddl} \\
  \hline
  latest supp. \pddl version      & considering recent \pddl features                     & 1.2                         & 3.1                       & 2.2                           &   3.1    &  3.1    & 3.1       \\
  syntax highlighting                 & supporting error detection and code navigation        & yes                         & basic                     & basic                         &   basic  & yes     & yes       \\
  semantic error detection            & supporting error detection                            & yes                         & no                        & no                            &   no     & yes     & no        \\
  automatic indentation               & supporting readability and navigation                 & no                          & no                        & yes                           &   no     & yes     & yes       \\
  code completion                     & speeding-up the knowledge engineering process         & yes                         & no                        & yes                           &   yes    & yes     & yes       \\
  code snippets                       & speeding-up the knowledge engineering process         & no                          & yes                       & yes                           &   basic  & yes     & yes       \\
                                      & externalizing user's memory                           &                             &                           &                               &          &         &           \\
  code folding                        & supporting keeping an overview of the code structure  & yes                         & no                        & yes                           &   yes    & yes     & yes       \\
  domain visualization                & supporting fast understanding of the domain structure & no                          & no                        & no                            &   no     & no      & yes       \\
  project management                  & supporting keeping an overview of associated files    & yes                         & yes                       & no                            &   yes    & yes     & yes       \\
  \uml to \pddl translation           & supporting initial modeling                           & no                          & yes                       & no                            &   no     & no      & no       \\
  planner integration                 & allowing for convenient planner access                & yes                       & yes                       & no                            &   yes    & yes     & yes       \\
  plan visualization                  & supporting understanding and crosschecking the plan   & no                          & yes                       & no                            &   no     & yes     & no        \\
  dynamic analysis                    & supporting dynamic domain analysis                    & no                          & yes                       & no                            &   no     & no      & no        \\
  declaration menu                    & supporting code navigation                            & no                          & no                        & yes                           &   no     & yes     & no        \\
  interface with                      & automating tasks                                      & no                          & no                        & no                            &   no     & no      & yes       \\
  programming language                & extending \pddl's modeling capabilities               &                             &                           &                               &          &         &           \\
  customization features              & acknowledging individual needs and preferences        & basic                       & no                        & yes                           &  basic   & yes     & yes       \\
\end{tabularx}
\label{tab:comparison}
\end{sidewaystable}

\section{MyPDDL}
\label{sec:mypddl}

\mypddl is a highly customizable and extensible modular system,
designed for supporting knowledge engineers in the process of writing,
analyzing and expanding \pddl files and thereby promoting the
collaboration between knowledge engineers and the use of \pddl in
real-world applications. The modules of \mypddl are described in the
next section.

\subsection{Modules}

\begin{description}
  \setlength\itemsep{1em}
\item[myPDDL-IDE] is an integrated development environment (IDE) for
  the use of \mypddl in the text and code editor \emph{Sublime
    Text}\footnote{\url{http://www.sublimetext.com}}. Since
  \mypddlsnippet and \textsc{-syntax} (see below) are devised
  explicitly for \sublimetext, their integration is implicit. The
  other tools described below (\mypddl-new, -diagram, -distance,
  -plan) can be used independently of \sublimetext via the
  command-line but can also be called from the editor.

\item[myPDDL-new] helps to organize \pddl projects. In many cases
  \pddl domains are created ad hoc \cite{shah2013exploring}. However,
  each implementation of a \pddl task specification comprises one
  domain and at least one corresponding problem file. Since several
  team members may be working on these files, keeping \pddl projects
  organized will facilitate collaboration. An automatically created,
  standardized project folder structure could facilitate the
  collaboration between users and the maintenance of consistency
  across projects. To this end, \mypddlnew creates the following
  folder structure when creating a new \pddl project:
  \begin{figure}[H]
  \dirtree{%
  .1 project-name/.
  .2 domains/.
  .2 problems/.
  .3 p01.pddl.
  .2 solutions/.
  .2 domain.pddl.
  .2 README.md.
  .2 plan.
  }
\end{figure}
All of the templates to create the files can be customized and new
templates can be added. The domain file \texttt{domain.pddl} and the
problem file \texttt{p01.pddl} initially contain corresponding \pddl
skeletons. Additionally the project name is used as the domain name
within the files \texttt{domain.pddl} and \texttt{p01.pddl}. All
problem files that are associated with one domain file are collected
in the folder \texttt{problems/}. \texttt{README.md} is a Markdown
file, which is intended for information about the authors of the
project, contact information, informal domain and problem
specifications, and licensing information.  Markdown files can be
converted to \textsc{html} by various hosting services like GitHub or
Bitbucket. The basic planner integration \mypddlplan provided by the
file \texttt{plan} is described below.

\item[myPDDL-snippet] provides code skeletons, that is, templates for
  often used \pddl constructs such as domains, problems, type and
  function declarations, and actions. They can be inserted by typing a
  triggering keyword. Table \ref{tab:snippets} displays descriptions
  of all available snippets and the corresponding trigger.

\begin{table}[htb]
  \centering
  \caption[Available snippets in \mypddl-snippet]{\label{tab:snippets}The snippets that can be inserted into \pddl files by typing the trigger.}
\begin{tabular}{ll}
  \tableheadline{snippet description} & \tableheadline{trigger}                   \\
  \hline
  domain skeleton                     & \texttt{domain}                           \\
  problem skeleton                    & \texttt{problem}                          \\
  type declaration                    & \texttt{t1, t2, ...}                      \\
  typed predicate declaration         & \texttt{p1, p2, ...}                     \\ 
  typed function declaration          & \texttt{f1, f2, ...}                     \\
  action skeleton                     & \texttt{action}, \texttt{durative-action} \\
\end{tabular}
\end{table}

For example, typing \texttt{action} and pressing the tabulator key
inserts a skeleton to specify an action. \pddl constructs with a
specified arity can be generated by adding the arity number to the
trigger (\texttt{p2} would insert the binary predicate template
\texttt{(pred-name~?x~-~object~?y~-~object)}).

Every snippet is stored in a separate file, located in the packages
folder of Sublime Text. New snippets can be added and existing
snippets can be customized by changing the templates in this folder.

\item[myPDDL-syntax] is a context-aware syntax highlighting feature
  for \sublimetext. It recognizes all \pddl constructs up to version
  3.1, such as comments, variables, names, and keywords and highlights
  them in different colors. Using regular expressions and a
  sophisticated pattern matching heuristic, it detects both the start
  and the end of \pddl code blocks and constructs. It then divides
  them into \emph{scopes}, that is, named regions. \sublimetext
  colorizes the code elements via the assigned scope names and in
  accordance with the current color scheme. These scopes allow for a
  fragmentation of the \pddl files, so that constructs are only
  highlighted if they appear in the correct context. Thus missing
  brackets, misplaced expressions and misspelled keywords are visually
  distinct and can be identified (Figure~\ref{fig:syntax}).

  \begin{figure}
    \centering
    \includegraphics[width=0.6\textwidth]{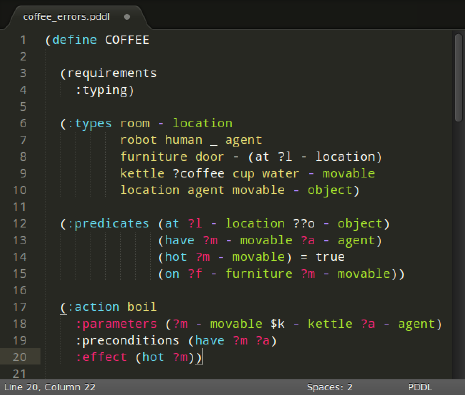}
    \caption{The figure shows the use of \mypddl in the text editor
      Sublime Text. Syntax errors in the domain are detected by
      \mypddlsyntax's context-aware syntax highlighting feature and
      displayed in white.}
\label{fig:syntax}
  \end{figure}

\item[myPDDL-clojure] provides a preprocessor for \pddl files to
  bypass \pddl's limited mathematical capabilities, thus reducing
  modeling time without overcharging planning algorithms. Since \pddl
  is used to create more and more complex domains
  \cite{goldman2012type,guerin2012academic}, one might need the square
  root function for a distance optimization problem or the logarithmic
  function for modeling an engineering problem. While these
  mathematical operations are currently not supported by \pddl itself,
  preprocessing \pddl files in a programming language and then
  hardcoding the results back into the file seem to be a reasonable
  workaround. With the help of such an interface, the modeling time
  can be reduced.  We decided to use the functional programming
  language Clojure \cite{hickey2008clojure}, a modern Lisp dialect,
  facilitating input and output of the Lisp-style \pddl
  constructs. Once a part is extracted and represented in Clojure, the
  processing possibilities are diverse and the full capacities of
  Clojure are available. It can be used for generating \pddl constructs,
  reading domain and problem files, handling, using and modifying the
  input, and generating \pddl files as output.

  The interface is provided as a Clojure library and based on two
  methods described below.
\begin{description}
\item[{read-construct(keyword, file)}] This methods allows for the
  extraction of code blocks from \pddl files. The following code block
  shows an example in which the goal state \texttt{(:goal (exploited
    magicfailureapp))} is extracted from a \pddl problem file.
\end{description}
Clojure command:
\begin{minted}[fontsize=\small,bgcolor=dhscodebg,rulecolor=\color{gray!40},frame=lines,framesep=5\fboxrule,framerule=1pt,tabsize=2]{clojure}
(read-construct :goal "garys-huge-problem.pddl")
;;=> ((:goal (exploited magicfailureapp)))
\end{minted}

\begin{description}
\item[{add-construct(file, block, part)}] This methods provides a
  means for adding constructs to a specified code block in \pddl
  domain and problem files. This is illustrated in the following two
  code blocks where the predicate \texttt{(hungry gisela)} is added to
  the \texttt{(:init~...)} block.
\end{description}
Clojure command:
\begin{minted}[fontsize=\small,bgcolor=dhscodebg,rulecolor=\color{gray!40},frame=lines,framesep=5\fboxrule,framerule=1pt,tabsize=2]{clojure}
(add-construct "garys-huge-problem.pddl" :init '((hungry gisela)))
\end{minted}
Updated \pddl file:
\begin{minted}[fontsize=\small,bgcolor=dhscodebg,rulecolor=\color{gray!40},frame=lines,framesep=5\fboxrule,framerule=1pt,tabsize=2]{text}
(:init (hungry gary)
       (in pizza-box big-pepperoni)
       (has-access gisela magicfailureapp))
       (hungry gisela))
\end{minted}

\item[myPDDL-distance] provides special preprocessing functions for
  distance calculations. In some domains, every object needs a
  location specified by $x$ and $y$ coordinates.  While the location
  of objects can be implemented using the predicate \texttt{(location
    ?o - object ?x ?y - number)}, with \texttt{x} and \texttt{y} being
  the spatial coordinates of an object, calculating the Euclidean
  distance requires using the square root function. However, \pddl 3.1
  supports only the four basic arithmetic operators.

  Parkinson and Longstaff~\cite{parkinson2012increasing} describe a
  workaround for this drawback. By writing an action
  \texttt{calculate-sqrt}, they bypass the missing square root
  function by making use of the Babylonian root method. Although this
  method approximates the square root function, it requires many
  iterations and would most likely have an adverse effect on plan
  generation \cite{parkinson2012increasing}.

  More usable and probably faster results can be achieved by using the
  interface between \pddl and Clojure as a distance calculator,
  implemented in the tool \mypddldistance. It reads a problem file
  into Clojure and extracts all locations, defined in the
  \texttt{(:init ...)} code block. The Euclidean distances between
  these locations are then calculated and written back into a new and
  now extended copy of the problem file, using the predicate
  \texttt{(distance~?o1~?o2~-~object ?n~-~number)}, which specifies
  the distance between two objects. The code blocks below show the
  \texttt{(:init ...)} block of a \pddl problem file before and after using
  \mypddl-distance.

  Before:
\begin{minted}[fontsize=\small,bgcolor=dhscodebg,rulecolor=\color{gray!40},frame=lines,framesep=5\fboxrule,framerule=1pt,tabsize=2]{text}
(:init ...
       (location gary 4 2)
       (location pizza 2 3))
     \end{minted}
     After:
\begin{minted}[fontsize=\small,bgcolor=dhscodebg,rulecolor=\color{gray!40},frame=lines,framesep=5\fboxrule,framerule=1pt,tabsize=2]{text}
(:init ...
       (location gary 4 2)
       (location pizza 2 3)
       (distance gary gary 0.0)
       (distance gary pizza 2.2361)
       (distance pizza gary 2.2361)
       (distance pizza pizza 0.0))
\end{minted}

The calculator works on any arity of the specified location predicate,
so that locations can be specified in 1D, 2D, 3D, and even used in
higher dimensions.

A disadvantage of this method is that the calculated distances have to
be stored in the \pddl problem file, potentially requiring many lines
of code. If the number of locations is \(n\), the number of calculated
distances is \(n^2\), so that every location has a distance to every
other location and itself. Therefore, a sensible next step would be
to extend \pddl by increasing its mathematical expressivity
\cite{parkinson2012increasing}, perhaps by declaring a requirement
\texttt{:math} that specifies further mathematical operations.

\item[myPDDL-diagram] generates a \textsc{png} image based on the type
  hierarchy of a \pddl domain file (Figure~\ref{fig:diagram}). The
  diagrammatic representation of textual information helps to quickly
  understand the connection of hierarchically structured items and
  should thus be able to simplify the communication and collaboration
  between developers. In the diagram, types are represented with
  boxes, with every box consisting of two parts:

  \begin{itemize}
    \setlength\itemsep{0.5em}
  \item The header displays the name of the type.
  \item The lower part displays all predicates that use the
    corresponding type at least once as a parameter. The predicates
    are written just as they appear in the \pddl code.
\end{itemize}

Generalization relationships express that every subtype is also an
instance of the illustrated super type (e.g. ``a hacker \emph{is a}
person). This relationship is indicated in the diagram with an
arrow from the subtype (here: \emph{hacker}) to the super type (here:
\emph{person}).

In order to create the diagram, \mypddl-diagram utilizes dot from the
Graphviz package~\cite{ellson2002graphviz} and takes the following
steps:

\begin{enumerate}
  \setlength\itemsep{0.5em}
\item A copy of the domain file is stored in the folder
  \texttt{domains/}.
\item The \texttt{(:types~...)} block is extracted via the PDDL/Clojure
  interface.
\item In Clojure, the types are split into super types and associated
  subtypes using regular expressions and stored in a Clojure hashmap.
\item Based on the hashmap, the description of a directed graph in the
  \textsc{dot} language is created and saved in the folder
  \texttt{dot/}.
\item The \textsc{dot} file is passed to dot, creating a \textsc{png}
  diagram and saving it in the folder
  \texttt{diagrams/}.
\item The \textsc{png} diagram is displayed in a window.
\end{enumerate}

Every time \mypddl-diagram is invoked, these steps are executed and,
optionally, the names of the saved files are extended by an ascending
revision number. Thus, one cannot only identify associated \pddl,
\textsc{dot} and \textsc{png} files, but also use this feature for
basic revision control.

  \begin{figure}
    \centering
    \includegraphics[width=1\textwidth]{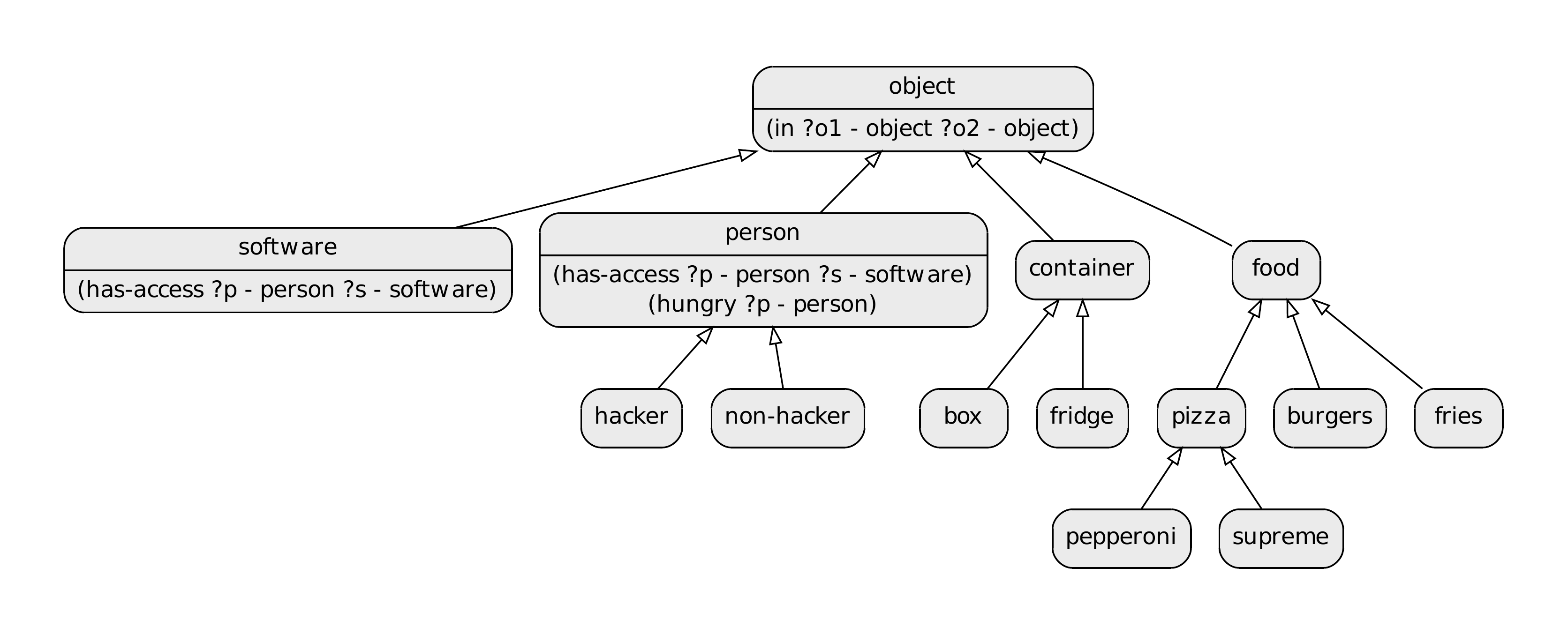}
    \caption{The type diagram generated by \mypddldiagram helps to
      grasp the relationship between types in the domain
      file. Additionally, it displays all predicates that use the
      corresponding type at least once as a parameter.}
    \label{fig:diagram}
  \end{figure}

\item[myPDDL-plan] is a basic planner integration for \mypddl. After
  creating a new project with \mypddlnew, the file \texttt{plan} in
  the project folder contains a shell script for executing a planner
  with the new domain and problem files as input. The desired planner
  can be specified in the file \texttt{plan} or by editing the
  templates of \mypddlnew. Due to the versatility of shell scripts,
  any planner can be used and arbitrary command line options can be
  specified. The planner can be invoked from Sublime Text or via the
  command line.
\end{description}

In order to provide easy installation and maintenance, \mypddlide can
be installed using \sublimetext's Package
Control\footnote{\url{https://sublime.wbond.net/about}}.  The project
source code is hosted on
GitHub\footnote{\url{https://github.com/Pold87/myPDDL}}, providing the
possibility to actively participate in the design
process. Additionally, \mypddlclojure is hosted on GitHub
\footnote{\url{https://github.com/Pold87/pddl-clojure-interface}} as
well as a standalone version to call the functions from a the command
line\footnote{\url{https://github.com/Pold87/pddl-clojure-interface-standalone}}.
The \mypddl project
site\footnote{\url{http://pold87.github.io/myPDDL/}} provides room for
discussing features and reporting bugs.

\section{Validation and Evaluation}
\label{sec:valid-eval}

To assess the utility of \mypddl, we evaluated its performance in
terms of collaboration, experience, efficiency, and debugging in a
user test. We analyzed the user performance both with and without
using \mypddlsyntax and \mypddldiagram.

\subsection{User Evaluation}

The two most central modules of \mypddl are \mypddlsyntax and
\mypddldiagram, since they support collaboration, efficiency, and
debugging independently of the user's experience with \pddl. To
evaluate their usability, they were evaluated in a user study. To this
end, we compared the user performance regarding several tasks, both
with and without using the respective module.

\subsubsection{Participants}

In Usability Engineering, a typical number of participants for user
tests is five to ten. Studies have shown that even such small sample
sizes identify about 80\,\% of the usability problems
\cite{nielsen1994estimating,hwang2010number}. Our study design
required eight participants. Three female and five male participants
took part in the study (average age of 22.9, standard deviation of age
0.6). All participants were required to have basic experience with at
least one Lisp dialect in order not to be confused with the many
parentheses, but no experience with \pddl or AI planning in general.

\subsubsection{Approach}

24 hours before the experiment was to take place, participants
received the web link\footnote{Tutorial in German:
  \url{https://www.youtube.com/watch?v=Uck-K8VnNOU&list=PL3CZzLUZuiIMWEfJxy-G6OxYVzUrvjwuV}}
to a 30-minute interactive video tutorial on AI planning and
\pddl. This method was chosen in order not to pressure the participant
with the presence of an experimenter when trying to understand the
material.

\subsubsection{Procedure}

We defined four tasks (Appendix~\ref{sub:tasks}): two debugging tasks
for testing the syntax highlighting feature and two type hierarchy
tasks for testing the type diagram generator. A within subjects design
was considered most suited due to the small number of
participants. Therefore, it was necessary to construct two tasks
matched in difficulty for each of these two types to compare the
effects of having the tools available. Each participant started either
with a debugging or type hierarchy task and was given the \mypddl
tools either in the first two tasks or the second two tasks, so that
each participant completed each task type once with and once without
\mypddl. This results in $2$ (first task is debugging or hierarchy)
$\times$ $2$ (task variations for debugging and hierarchy) $\times$
$2$ (starting with or without \mypddl) $= 8$ individual task orders,
one per participant.

\begin{itemize}
\item Debugging Tasks

  For the debugging tasks, participants were given six minutes (a
  reasonable time frame tested on two pilot tests) to detect as many
  of the errors in the given domain as possible. They were asked to
  record each error in a table using pen and paper with the line
  number and a short comment. Moreover, they were instructed to
  immediately correct the errors in the code if they knew how to, but
  not to dwell on the correction otherwise. For the type hierarchy
  task, participants were asked to answer five questions concerning
  the domains, all of which could be facilitated with the type diagram
  generator. One of the five questions (Question 4, see
  Appendix~\ref{splisus} and \ref{store}) also required looking into
  the code. Participants were told that they should not feel pressured
  to answer quickly, but to not waste time either. Also they were
  asked to say their answer out loud as soon as it became evident to
  them. They were not told that the time it took them to come up with
  an answer was recorded, since this could have made them feel
  pressured and thus led to more false answers.

\item Type Hierarchy Tasks

  The two tasks to test syntax highlighting presented the user with
  domains that were 54 lines in length, consisted of 1605 characters
  and contained 17 errors each. Errors were distributed evenly
  throughout the domains and were categorized into different
  types. The occurrence frequencies of these types were matched across
  domains as well, to ensure equal difficulty for both domains. To
  test the type diagram generator, two fictional domains with equally
  complex type hierarchies consisting of non-words were designed (five
  and six layers in depth, 20 and 21 types). The domains were also
  matched in length and overall complexity: five and six predicates
  with approximately the same distribution of arities, one action with
  four predicates in the precondition and two and three predicates in
  the effect.

\item System Usability Scale

  At the end of the usability test the participants were asked to
  evaluate the perceived usability of \mypddl using the system
  usability scale \cite{brooke1996sus}.

\end{itemize}

\newpage
\subsubsection{Analysis}
\begin{itemize}
\item Debugging Tasks: To compare differences in the
  debugging tasks, a paired sample $t$-test was used; normality was
  tested with a Shapiro-Wilk test. To compare the arithmetic means
  ($M$s) of detected errors, the test was performed two-tailed, since
  syntax highlighting might both help or hinder the
  participants. Arithmetic standard deviations ($SD$s) were calculated
  for each condition.

\item Type Hierarchy Tasks:
\nopagebreak
  For the type hierarchy tasks, $t$-tests were performed on the
  logarithms of the data values to compare the geometric means for the
  two conditions for each question; normality was tested with a
  Shapiro-Wilk test on the log-normalized data values. The geometric
  mean is a more accurate measure of the mean for small sample sizes
  as task times have a strong tendency to be positively skewed
  \cite{sauro2012quantifying}. The geometric standard deviation
  ($GSD$) was calculated for each question and condition. Only those
  task completion times were included in the calculation of the
  $t$-values, where the respective participant gave a correct answer
  for both occurrences of a question. This approach should reduce the
  influence of random guessing. Again, two-tailed $t$-tests were used
  to account for both, improvements and drawbacks, of using
  \mypddldiagram.

\item System Usability Scale

  The arithmetic mean and standard deviation for the score on the
  System Usability Scale was calculated.
\end{itemize}

\subsubsection{Results}

\begin{itemize}
\item Debugging Tasks

  The participants detected more errors using the syntax highlighting
  feature ($M = 10.3$, $SD = 3.45$) than without it ($M = 7.6$,
  $SD = 2.07$); $t(7) = 2.68$, $p = 0.03$.  That is, approximately
  36\,\% more errors were found with syntax highlighting. The
  arithmetic means are displayed in Figure
  \ref{fig:found-errors-combined}, where each cross~($\times$)
  represents the data value of one participant.
\begin{figure}[h]
  \centering
  \hspace{1.7cm}
  \includegraphics[width=0.7\textwidth]{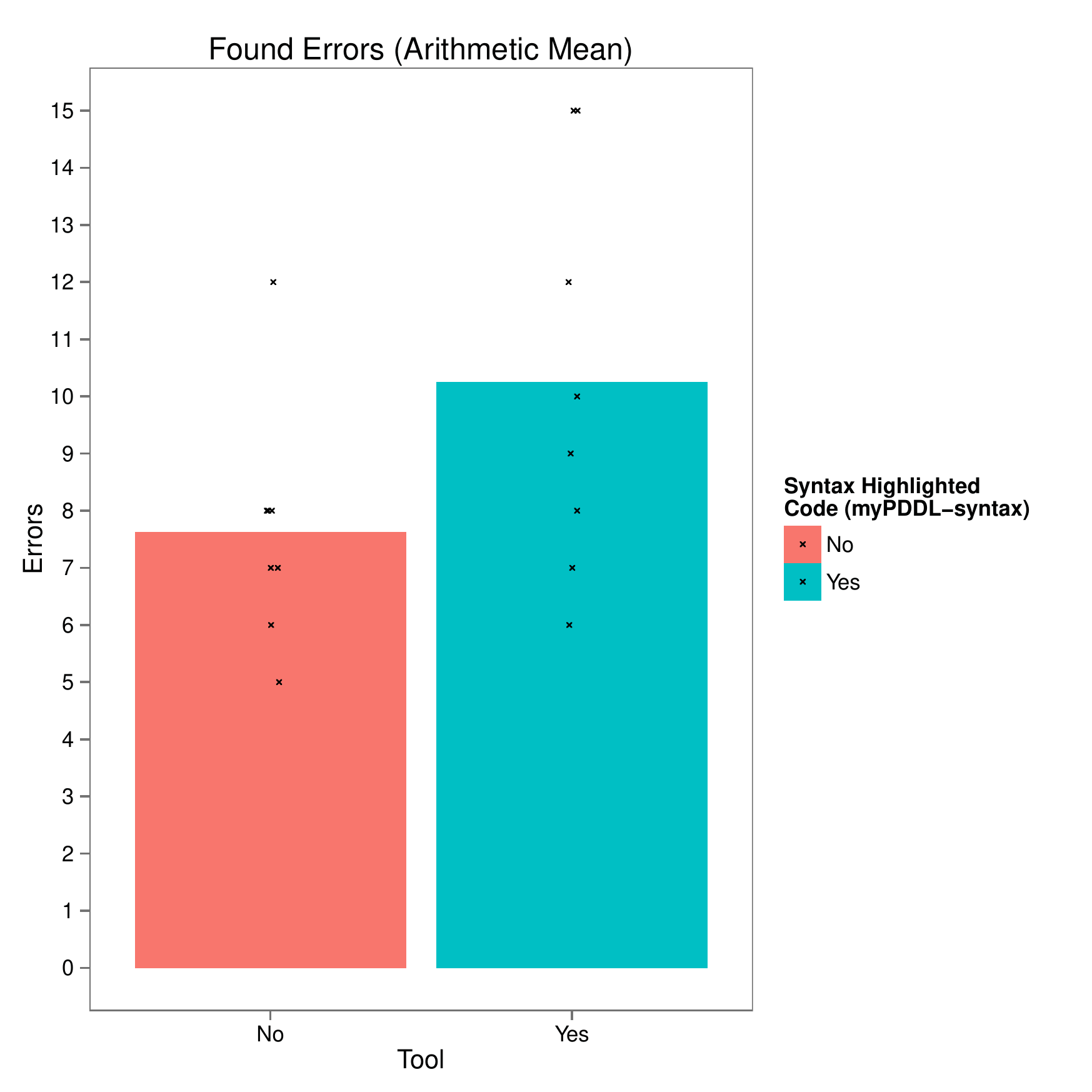}
  \caption[Diagram of detected errors]{Comparison of detected errors
    with and without the syntax highlighting feature. Each cross
    ($\times$) shows the data value of one participant. The bars
    display the arithmetic mean.}
\label{fig:found-errors-combined}
\end{figure}

\item Type Hierarchy Tasks

  Figure~\ref{fig:task-completions-agg} shows the geometric mean of
  the completion time of successful tasks for each question with and
  without the type diagram generator. With the type diagram generator
  participants answered all questions (except Question 4) on average
  nearly twice as fast ($GM = 33.0$, $GSD = 2.23$) as without it
  ($GM = 57.8$, $GSD = 2.05$); $t(32) = -3.34$, $p = 0.002$. This
  difference slightly increases, if Question 4 is excluded from the
  calculations: with type generator: $GM = 31.1$, $GSD = 2.17$,
  without: $GM = 58.1$, $GSD = 2.07$; $t(30) = -3.68$, $p < 0.001$.
  Table~\ref{tab:p-values} gives an overview of geometric means,
  geometric standard deviations, $t$-values, and $p$-values for each
  question.

  \begin{table}[]
    \centering
    \begin{tabular}{rrrrrrrr}
    \toprule
         & \multicolumn{4}{c}{Type Diagram Generator} & & &                                                                   \\
         & \multicolumn{2}{c}{with}                   & \multicolumn{2}{c}{without} &      &       &                     \\
\cmidrule(r){2-3}\cmidrule(r){4-5}
Question & $GM$                                       & $GSD$                       & $GM$ & $GSD$ & $df$ & $t$   & $p$  \\
\midrule
      Q1 & 21.8                                       & 1.52                        & 40.0 & 2.26  & 7    & -1.86 & 0.11 \\
      Q2 & 23.8                                       & 1.49                        & 50.8 & 2.16  & 7    & -1.91 & 0.10 \\
      Q3 & 48.0                                       & 3.49                        & 83.2 & 2.20  & 5    & -0.86 & 0.43 \\
      Q4 & 84.3                                       & 2.22                        & 54.1 & 1.93  & 1    & 4.48  & 0.14 \\
      Q5 & 41.2                                       & 2.24                        & 78.0 & 1.48  & 7    & -2.75 & 0.03 \\
\bottomrule
    \end{tabular}
    \caption{Overview of geometric means ($GM$s), geometric standard deviations ($GSD$s), degrees of freedom ($df$), $t$-values, and $p$-values. The calculation for Q4 is based on only two paired data values ($df = 1$). This table only considers paired data values, this means only if a participant answered the question correctly in both domains, the data value is considered (since \emph{paired} $t$-tests are calculated). In contrast, Figure \ref{fig:task-completions-agg} displays the geometric means for all correct answers.}
    \label{tab:p-values}
  \end{table}

\begin{figure}[h]
  \centering
  \hspace{1.7cm}
  \includegraphics[width=0.85\textwidth]{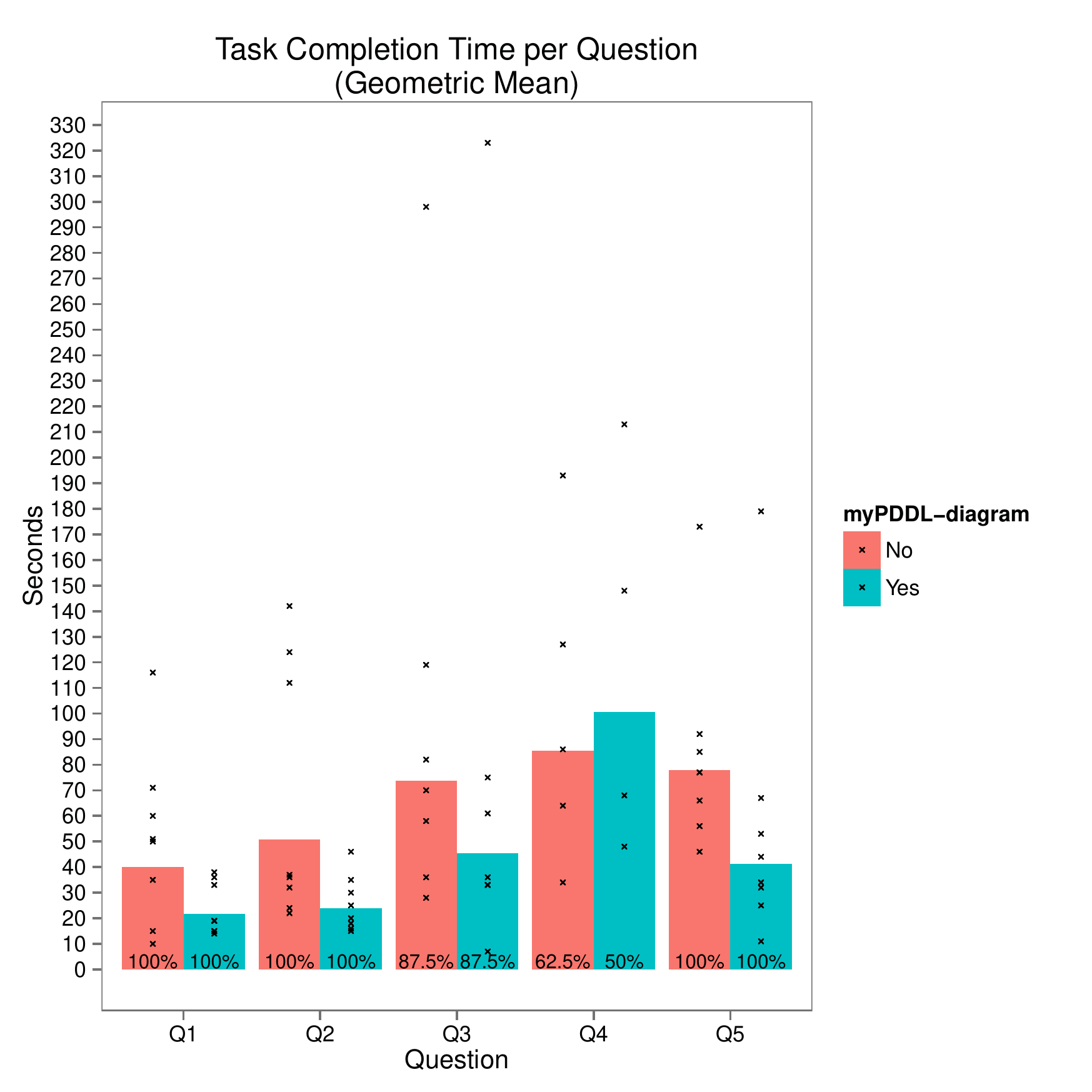}
  \caption[Diagram of the task completion time]{Task completion time
    for the type hierarchy tasks. The bars display the geometric mean
    averaged over all participants; each cross~($\times$) represents
    the data value of one participant. The percent values at the
    bottom of the bars show the percentage of users that completed the
    task successfully. The questions can be found in the
    Appendix~\ref{splisus} and
    \ref{store}.}\label{fig:task-completions-agg}
\end{figure}

\item System Usability Scale
  \mypddl reached a score of 89.6 on the system usability scale
  \cite{brooke1996sus}, with a standard deviation of 3.9. 
\end{itemize}

\subsubsection{Discussion}

The user test shows that \mypddlsyntax and -\textsc{diagram} provide useful
tools for novices in AI planning and \pddl. Below, we will discuss
each part of the user test in turn.

\begin{itemize}
\item Debugging Tasks

  While, in general, the syntax highlighting feature was considered
  very useful, two participants remarked that the used colors confused
  them and that they found them more distracting than helpful. One of
  them mentioned that the contrast of the colors was so low that they
  were hard for her to distinguish. She found the same number of
  errors with and without syntax highlighting. The other of the two
  was the only participant who found less errors with syntax
  highlighting than without it. With \mypddlsyntax, two participants
  found all errors in the domain, while none achieved this without
  syntax highlighting. While every participant had to use the same
  color scheme in the experiment, colors are customizable in
  \sublimetext.

\item Type Hierarchy Tasks

  In spite of the rather large difference between the $GM$s for
  Question 3, a high $p$-value is obtained ($p = 0.43$). This might be
  due to the high $GSD$ for the \emph{with} condition and the rather
  small degrees of freedom ($df = 5$). Testing more participants would
  probably yield clearer results here. The fact that the availability
  of tools did not have a positive effect on task completion times for
  Question 4 can probably be attributed to the complexity of this
  question (see Appendix~\ref{splisus} and \ref{store}): in contrast
  to the other four questions, here, participants were required to
  look at the actions in the domain file in addition to the type
  diagram. Most participants were confused by this, because they had
  assumed that once having the type diagram available, it alone would
  suffice to answer all questions. This initial confusion cost some
  time, thus negatively influencing the time on the task.

  Visualization tools such as \mypddl -diagram can improve the
  understanding of unknown \pddl code and thus support
  collaboration. But users may be unaware of the limitations of such
  tools. A possible solution is to extend \mypddl -diagram to display
  actions, but this can overload the diagram and, especially for large
  domains, render it unreadable. Different views for different aspects
  of the domain or dynamically displayed content could integrate more
  data, but this also hides functionality, which is generally
  undesired for usability \cite{norman2002design}.

\item Sytem Usability Scale

Since the overall mean score of the system usability scale has an
approximate value of 68 with a standard deviation of 12.5
\cite{sauro2011practical}, the score of \mypddl is well above average
with a small standard deviation. A score of 89.6 is usually attributed
to superior products \cite{bangor2008empirical}. Furthermore, 89.6
corresponds approximately to a percentile rank of 99.8\,\%, meaning
that it has a better perceived ease-of-use than 99.8\,\% of the
products in the database used by Sauro \cite{sauro2011practical}.

\end{itemize}

In summary, the user test shows that customizability is important, as
not all users prefer the same colors or syntax highlighting at all and
their personal preferences seem to correlate with the effectiveness of
the tools.

\section{Conclusion}
\label{sec:conclusion}

We designed \mypddl to support knowledge engineers in creating,
understanding, modifying, and extending planning domains. \mypddl's
code editing features such as syntax highlighting and code snippets,
as well as a type diagram generator, an interface with the programming
language Clojure, and a planner integration can help in the various
stages of working with \pddl domains. \mypddl's extensible and
customizable architecture helps to fulfill the different preferences
and requirements of knowledge engineers. In the conducted user test,
\mypddl users were able to grasp the domain structure of a \pddl file
more quickly than non-users and also found more errors in a
deliberately erroneous domain file. Moreover, the users found the
tools easy and pleasant to use.

In future work, \mypddl's set of features could be extended in several
directions. The interface between \pddl and Clojure offers a basis for
creating dynamic planning scenarios. Applications could be the
modeling of learning and forgetting by adding facts to or retracting
facts from a \pddl file or the modeling of an ever changing real world
via dynamic predicate lists. Another way of putting the interface to
use would be by making the planning process more interactive, allowing
for the online interception of planning software in order to account
for the needs and wishes of the end user. Since many features of
\mypddl can be called via the command line, interfaces with other
editors could be developed. So far, there is a basic integration with
the code editor
Atom\footnote{\url{https://github.com/Pold87/myPDDL-Atom}}.

All in all, the overall increase of efficiency due to facilitated
collaboration and support in maintaining an overview should encourage
a shift of focus toward real world problems in knowledge
engineering. The full modeling potential can only be reached with
appropriate tools, with \mypddl hopefully leading to a broader
acceptance and use of \pddl for planning problems.

\bibliographystyle{splncs04}
\bibliography{mybibliography}
\newpage
\appendix

\section{Tasks}
\label{sub:tasks}
\subsection{Deliberately erroneous Logistics Domain}
\label{logistics}
\begin{minted}[fontsize=\tiny,bgcolor=dhscodebg,rulecolor=\color{gray!40},frame=lines,framesep=5\fboxrule,framerule=1pt,tabsize=2]{text}
;;;; Logistics domain

(define (domain ?logistics)

  (:requirements
    :types) 

  (:typing truck airplane motorboat - vehicle
           package vehicle suitcase furniture - thing
           airport garage station - location
           car1 car 2 car3 - vehicle
           city location thing - object)

  (:predicates (in-city ?l - location ?c - city)
               (at ?obj - thing ?l - location)
               (key ?v - vehicle) = true
               (full ?v - vehicle)
               (in ?p - package ??veh - vehicle))

  (:action drive
    :parameters (?t - truck ?from ?to - location ?c - city)
    :precondition (and (at ?tr ?from)
                       (in-city ?from ?c)
                       (incity ?to ?c))
    :effect (and (not (at ?t ?from))
                 (at ?t ?to)))

  (:action fly
    :parameters (?a - airplane ?from ?to - airport)
    :precondition (at ?a ?from)
    :effect (and (n0t (at ?a ?from))
                 (at ?a ?to)))

  (:action fuel
    :parameters (?v - vehicle ?c - city ?to airport)
    :precondition (and (not (full ?v))
                       (in-city ?to ?c)
                       (at ?v ?to))
    :effect (full ?v))                   

  (:action load
    parameters: (?v - vehicle ?p - package ?l - location)
    precondition: (and (?v ?l)
                       (at ?p ?l))
    :effect (and (ay ?p ?l)
                 (in ?p ?v)))

  (:action unload
    :parameters (?v - vehicle p - package ?l - location)
    :precondition (and (at ?v ?l)
                           ?p ?v)
    :effects (and (not (in ?p ?v))
                  (at ?p - ?l))))
                \end{minted}
The original file can be downloaded at
\url{http://ipc.informatik.uni-freiburg.de/PddlExtension}:

\subsection{Deliberately erroneous Coffee Domain}
\label{coffee}
\begin{minted}[fontsize=\scriptsize,bgcolor=dhscodebg,rulecolor=\color{gray!40},frame=lines,framesep=5\fboxrule,framerule=1pt,tabsize=2]{text}
(define COFFEE

  (requirements
    :typing)

  (:types room - location
                 robot human _ agent
                 furniture door - (at ?l - location)
                 kettle ?coffee cup water - movable
                 location agent movable - object)

  (:predicates (at ?l - location ??o - object)
               (have ?m - movable ?a - agent)
               (hot ?m - movable) = true
               (on ?f - furniture ?m - movable))

  (:action boil
    :parameters (?m - movable \$k - kettle ?a - agent)
    :preconditions (have ?m ?a)
    :effect (hot ?m))

  (:action grip-some
    :parameters (?m - movable ?r - robot ?f - _furniture ?l - location)
    :precondition (and (at ?l ?r)
                       (on ?fu ?m)
                       (at ?l ?f))
    :effect (and (have ?m ?r)))

  (:action move
    :parameters: (?m - movable ?a - agent ?from ?to - location)
    :precondition (or (änd (at ?from ?a)
                           (at ?from ?m))
                           (and (at ?from ?m)
                                (location ?from ?a)))
    :effect (and (not (at ?from ?m))
                 (at ?to ?m)))

  (:action change-room
    :parameters (?from-r ?to-r - room ?a - agent)
    :precondition (at ?fromr ?a) 
    :effect (and (not (at ?from-r ?a))
                 (at ?tor ?a)))

  (:action prep-coffee
    :parameters (?a - agent ?c - cjp ?w - water ?cof - coffee)
    :precondition (and (have ?c ?a)
                       (hot ?w))
    :effect (have ?cof ?a))

  (:action ?hand-over
    :parameters (?m - movable ?a1 - agent ?a2 - agent)
    :precondition (have ?m ?a1))
    :effect (and (not (have ?m ?a1))
                 (have ?m ?a2))))
\end{minted}

\subsection{Planet Splisus}
\label{splisus}
\begin{minted}[fontsize=\scriptsize,bgcolor=dhscodebg,rulecolor=\color{gray!40},frame=lines,framesep=5\fboxrule,framerule=1pt,tabsize=2]{text}
(define (domain splisus) 

  (:requirements :typing)

  (:types splis - gid
          spleus - splos
          schprok schlok - splus
          rud mekle - lech
          hulpf hurpf - hupf
          sipsi flipsi hupf - splis
          schmok schkok - splus
          gid splos splus - ruffisplisus
          merle - hupf
          ruffisplisus mak lech - object)

  (:predicates (father-of ?r1 - ruffisplisus ?r2 - ruffisplisus)
               (married ?s1 - splos ?s2 - splis)
               (has-weapon ?h - sipsi)
               (dead ?r1 - ruffisplisus)
               (at ?l - lech ?r - ruffisplisus))

  (:action kill
    :parameters (?l - lech ?r1 - ruffisplisus ?s - splis)
    :precondition (and (at ?l ?r1)
                       (at ?l ?s)
                       (married ?r1 ?s)
                       (has-weapon ?s))
    :effect (and (dead ?r1)
            (not (married ?r1 ?s)))))
\end{minted}

Please answer the following five questions on the society and
structure of Planet Splisus:
\begin{enumerate}
\item Are all \emph{Flipsis} also of the type \emph{Ruffisplisus}?
\item Are all \emph{Merles} also \emph{Splus}?
\item Can a \emph{Spleus} be married to a \emph{Schlok}?
\item Only theoretically: Could a \emph{Hurpf} murder a \emph{Spleus}?
\item Let us assume there are three categories of object types on
  Splisus: places, beings and food. Match the three object types
  \emph{Ruffisplisus}, \emph{Mak} and \emph{Lech} with these
  categories.
\end{enumerate}

\subsection{Store}
\label{store}
\begin{minted}[fontsize=\scriptsize,bgcolor=dhscodebg,rulecolor=\color{gray!40},frame=lines,framesep=5\fboxrule,framerule=1pt,tabsize=2]{text}
(define (domain store)

  (:requirements :typing)

  (:types lala lila - zahls
          blisis blusis - ultri
          iltre lula - nulls
          zahls schwinds - knozi
          minis - lala
          ultri sopple schmitzl - lila
          ultres raglos wexis - lola
          kosta - nulls
          nulls spax - minis
          lola - zahls
          knozi schmus - object)

  (:predicates (product ?k - knozi) ; Produkt
               (workplace ?l1 - lola ?l2 - lala) 
               (product-at ?l1 - lola ?l2 - lila) 
               (cashier ?k - knozi) 
               (customer ?s - spax)
               (owns ?l - lila ?s - spax)) 

  (:action sell
    :parameters (?p - lila ?z - zahls ?l - lola ?w - wexis ?s - spax)
    :precondition (and (product ?p)
                       (cashier ?z)
                       (product-at ?l ?p)
                       (customer ?s))
    :effect (and (product-at ?w ?p)
                 (not (product-at ?l ?p))
                 (owns ?p ?s))))
\end{minted}
Please answer the following five questions concerning the environment
store:
\begin{enumerate}
\item Are objects of the type \emph{Lula} also of the type
    \emph{Minis}?
\item Are \emph{Spax} and \emph{Schmus} \emph{Zahls}?
\item Is it possible for an Iltre to work at a
    \emph{workplace} of the type \emph{Knozi}?
\item Only theoretically: Could a \emph{Lala} sell a
    \emph{Schmitzl} to a \emph{Kosta}?
\item Let us assume our domain \emph{store} models a
    grocery store. There are three categories: humans, products, and
    places. Can you match these world terms with the object types
    \emph{lila}, \emph{lala}, and \emph{lola} from the domain?
\end{enumerate}

\end{document}